\documentclass[a4paper,11pt]{article}
\usepackage{amssymb,amsmath,bm}  
\usepackage{graphics}   
\graphicspath{ {figures/} }
\usepackage{array,booktabs}      
\usepackage{authblk}  

\usepackage[left= 2.5cm, right=2.5cm, top=2.5cm]{geometry}

\usepackage[pdftex]{graphicx}

\usepackage{color}
\definecolor{coolblack}{rgb}{0.0, 0.18, 0.39}

\usepackage{cite}

\usepackage[utf8]{inputenc}
\usepackage{makecell}

\usepackage[linktocpage,
colorlinks = true,
linkcolor = blue,
urlcolor  = coolblack,
citecolor = red,
anchorcolor = blue]{hyperref}

\usepackage{epsfig}

\numberwithin{equation}{section}

\title{\textbf{Accidental scale-invariant Majorana dark matter
in leptoquark-Higgs portals}}
\author[1]{Ahmad Mohamadnejad\thanks{a.mohamadnejad@ut.ac.ir}}
\affil[1]{Young Researchers and Elite Club, Islamshahr Branch, Islamic Azad University, Islamshahr 3314767653, Iran}
\date{\today}

\begin{document}

\baselineskip 0.65 cm

\maketitle

\begin{abstract}
We study a classically accidental scale-invariant extension of the Standard Model (SM) containing three additional fields, a vector leptoquark ($ V_{\mu} $), a real scalar ($ \phi $), and a neutral Majorana fermion ($ \chi $) as a dark matter (DM) candidate. The scalar $ \phi $ (scalon) and Majorana fermion $ \chi $ are both singlets under the SM gauge group, while $ V_{\mu} $ has (\textbf{3}, \textbf{1}, 2/3) quantum numbers under the
$ SU(3)_{c} \times SU(2)_{L} \times U(1)_{Y} $.
The Majorana DM couples to the SM sector via both Higgs and leptoquark portals.
We perform a scan over the independent parameters
to determine the viable parameter space consistent with the Planck data for DM relic density, and
with the PandaX-II and LUX direct detection limits for the spin-independent (SI) and spin-dependent (SD) DM-nucleon cross section. The model generally evades indirect detection constraints while being consistent with collider data.
\end{abstract}



\section{Introduction}
Cosmological observations implies that DM is the majority of matter in the Universe. It is not made of SM particles and understanding its nature is one of the most important issues at the frontier of particle physics \cite{Bertone:2004pz}.

On the other hand, SM is expected to be valid up to energies of the order
of the Planck scale where vacuum stability problem arises. One solution to this problem is the supersymmetric extensions of SM where the Higgs
mass is radiatively stable down to the scale of supersymmetry breaking. However, the results from LHC have been negative for supersymmetry so far. Another solution is scale-invariant extensions of SM with no dimensionful parameter \cite{Bardeen}. In scale-invariant extensions of SM, all physical masses arise via Coleman-Weinberg mechanism \cite{Coleman:1973jx}.
This mechanism works only if extra bosonic degrees of freedom are added to SM with sizable couplings. Scale-invariant extensions of SM are also a generic feature of many DM models with bosonic \cite{Foot:2010av,Ishiwata:2011aa,Gabrielli:2013hma,Hambye:2013dgv,Carone:2013wla,Khoze:2014xha,Guo:2014bha,Endo:2015nba,
Wang:2015cda,Ghorbani:2015xvz,Plascencia:2015xwa,Karam:2015jta,Karam:2016rsz,Khoze:2016zfi,YaserAyazi:2019caf,Jung:2019dog} and fermionic \cite{Radovcic:2014rea,Altmannshofer:2014vra,Benic:2014aga,Ahriche:2015loa,Ahriche:2016ixu,Oda:2017kwl,YaserAyazi:2018lrv} DM candidates.

In this paper, we study an accidental scale-invariant extension of SM with vector leptoquarks as extra bosonic degrees of freedom mediating lepton-quark interactions. Leptoquarks are a natural result of unification of quarks and leptons \cite{Dorsner:2016wpm} initially proposed in the Pati-Salam model \cite{Pati:1974yy}. Leptoquarks would turn leptons into quarks generating new physical effects.
Leptoquarks also appear in SUSY models with R-parity violation \cite{Dreiner:1991pe,Butterworth:1992tc,Romao:1992vu}, and in composite models of leptons and quarks \cite{Schrempp:1984nj}. 
Besides vector leptoquarks, we also introduce a Majorana DM candidate and a real scalar field which is needed in order to get mass term for Majorana DM after symmetry breaking. In this model, DM mediates with SM via Higgs and leptoquark portals.

Leptoquarks can explain
some deviations from the SM such as anomalous B decays observed in BaBar \cite{Lees:2012xj,Lees:2013uzd}, Belle \cite{Huschle:2015rga} and LHCb \cite{Aaij:2015yra,Aubert:2007dsa,Bozek:2010xy}, a violation in lepton universality \cite{Aaij:2014ora} and a deviation from the SM prediction of $ (g-2)_{\mu} $ \cite{Davier:2010nc,Queiroz:2014zfa}. It is also shown that all three anomalies could be interpreted via the addition of a single scalar leptoquark \cite{Bauer:2015knc}.
DM models with scalar leptoquark portal can be found
in \cite{Agrawal:2011ze,Allahverdi:2017edd,Fornal:2018eol,Garny:2018icg,Queiroz:2014pra}. Vector leptoquark portal is also studied in \cite{Mandal:2018czf}. In our scenario, vector leptoquark is not a gauge field, however, it gets mass via its coupling to scalar fields. Particularly, we study the case in which vector leptoquark couples to scalon and the
spontaneous symmetry breaking makes it massive. Some attempts to write a model with gauge leptoquarks can be found in \cite{Assad:2017iib,DiLuzio:2017vat,Calibbi:2017qbu}.
Lately, the vector leptoquark has
also been considered as a possible explanation of the anomalies observed in charged-current and neutral current transitions of B mesons \cite{Barbieri:2015yvd,DiLuzio:2017chi,Choudhury:2017qyt,Blanke:2018sro,Crivellin:2018yvo}.

Majorana DM can leave detectable signals
at direct detection experiments. Both spin-independent (SI) and spin-dependent (SD) DM-nucleon scattering occur in our model. It is because Majorana DM interacts with SM via Higgs portal (with SI DM-nucleon scattering) as well as vector leptoquark portal (with SD DM-nucleon scattering). Hence, our model provide more opportunity to be probed compared to
a Majorana fermion DM with one portal either Higgs or vector leptoquark. The PICO \cite{Amole:2019fdf} and LUX \cite{Akerib:2016lao} data for DM-nucleon SD
 cross section allows the region compatible with
relic density and does not constrain the model. However, 
the direct detection experiments such as XENON1T \cite{Aprile:2017iyp,Aprile:2018dbl}, LUX \cite{Akerib:2016vxi}, and PandaX-II \cite{Cui:2017nnn} impose bounds on the SI DM-nuclei cross section. We also show that indirect detection experiments such as Fermi Large Area Telescope (Fermi-LAT) \cite{Ackermann:2015zua} and Alpha Magnetic Spectrometer (AMS) \cite{Aguilar:2016kjl} do not constrain our model. Finally, our model is compatible with collider physics.

The paper is organized as follows. In section \ref{The model}, we introduce the model. Section \ref{Relic density} contains
the calculation of DM relic density. In section \ref{Direct detection}, the SI and SD DM-nucleon cross section for direct detection experiments as well as DM indirect detection are studied. Finally, our conclusion including a discussion on recent collider bounds comes in section \ref{Conclusion}.

\section{The model} \label{The model}
We begin with constructing a model in which all couplings are diemnsionless. The fields gain mass via radiative Coleman-Weinberg symmetry breaking at one-loop level \cite{Coleman:1973jx}. 
Therefore, the model is a scale-invariant extension of SM without Higgs mass term. 

Apart from SM fields, the model contains three new fields which two of them are singlets under SM gauge transformation. These two fields are the real scalar $ \phi $ and the Majorana spinor $ \chi $. 
The other field, $ V_{\mu} $, is a vector leptoquark which has (\textbf{3}, \textbf{1}, 2/3) quantum numbers under the
$ SU(3)_{c} \times SU(2)_{L} \times U(1)_{Y} $ gauge group. This vector leptoquark does not lead to proton decay and it is also a part of the gauge sector of the Pati-Salam model \cite{Pati:1974yy}. 

Putting together these fields, and regarding scale invariance, gauge invariance, and renormalization conditions, we get
\begin{align} \label{2-1}
{\cal L} \supset& \, \, \frac{1}{2} \partial_{\mu} \phi \, \partial^{\mu} \phi  +  \frac{1}{2} i \overline{\chi} \gamma^{\mu} \partial_{\mu} \chi  - \frac{1}{2} V_{\mu \nu}^{\dagger} V^{\mu \nu}
   \nonumber \\
& - \frac{1}{6} \lambda_{H} (H^{\dagger}H)^{2}
- \frac{1}{4!} \lambda_{\phi} \phi^{4}
 - \lambda_{\phi H} \phi^{2}  H^{\dagger}H \nonumber \\
& - \lambda_{H V}  H^{\dagger}H V_{\mu}^{\dagger} V^{\mu} - \lambda_{ \phi V} \phi^{2}  V_{\mu}^{\dagger} V^{\mu} - \frac{1}{2} g_{\phi} \phi \overline{\chi}  \chi  \nonumber \\
 & - \sum_{\text{generations}} ( g_{L} \overline{q}_{L} \gamma_{\mu} V^{\mu} l_{L} + 
 g_{R}  \overline{d}_{R} \gamma_{\mu} V^{\mu} l_{R} 
 + g_{\chi}  \overline{u}_{R} \gamma_{\mu} V^{\mu} \chi  + \text{h.c.}),
\end{align}
where
\begin{align} \label{2-2}
& V_{\mu \nu} = D_{\mu} V_{\nu} - D_{\nu} V_{\mu} ,
   \nonumber \\
& D_{\mu} = \partial_{\mu} - i g_{s} \frac{\lambda^{a}}{2} G^{a}_{\mu} - i g_{Y} Y  B_{\mu} .
\end{align}
In DM models with leptoquark portal, we do not necessarily need the real scalar field, however in our scenario, because of scale invariance condition, we need this field in order to get mass term for Majorana spinor after symmetry breaking. Therefore, DM interacts with SM particles via both Higgs and vector leptoquark portals.

Note that the model can not be fundamental and might be regarded as an effective theory. As we mentioned in Introduction, $ V_{\mu} $ is not a gauge field, i.e., Lagrangian (\ref{2-1}) is not invariant under gauge transformation $ V_{\mu} \rightarrow U V_{\mu} U^{\dagger} - \frac{i}{g} (\partial_{\mu} U) U^{\dagger} $ where $ U $ presents some gauge group. Instead it is invariant under $ V_{\mu} \rightarrow U V_{\mu} U^{\dagger}$ where for $ U $ being SM gauge group, symmetry properties of $ V_{\mu} $ is mentioned in table~\ref{table1}. Nonetheless, there will be an accidental gauge symmetry when neglecting all the interactions. In this case, the internal degrees of freedom (dof) of $ V_{\mu} $ is $ 2 \times 3 \times 2 = 12 $, where factor 2 is for charge, 3 for color, and the other 2 for transverse degrees of freedom. After symmetry breaking, leptoquarks become massive and dof will be $ 2 \times 3 \times 3 = 18 $ (the last 3 is for transverse and longitudinal degrees of freedom). Therefore, dof would not match. To avoid this problem, one should consider a small mass term for vector leptoquark before the "scalon" gets a VEV. It means the scale-invariant symmetry in \eqref{2-1} is approximate and the theory should be considered as an accidental scale-invariant model.

\begin{table}
\centering
\begin{tabular}{c c c} 
 \hline
 Field & Symbol & $ (SU(3)_{c} , SU(2)_{L} , U(1)_{Y}) $ \\ [0.5ex] 
 \hline
Scalon & $ \phi $ & $(\textbf{1},\textbf{1},0)$ \\
Higgs doublet & $ H $  & $(\textbf{1},\textbf{2},\frac{1}{2}) $ \\
Left-handed leptons & $ l_{L} $  & $ (\textbf{1},\textbf{2},-\frac{1}{2}) $ \\
Right-handed leptons & $ l_{R} $  & $(\textbf{1},\textbf{1},-1)$ \\
Left-handed quarks & $ q_{L} $  & $ (\textbf{3},\textbf{2}, \frac{1}{6}) $ \\
Right-handed quarks (up-type) & $ u_{R} $  & $ (\textbf{3},\textbf{1}, \frac{2}{3}) $ \\
Right-handed quarks (down-type) & $ d_{R} $  & $(\textbf{3},\textbf{1},-\frac{1}{3} ) $ \\
Majorana DM & $ \chi $ & $(\textbf{1},\textbf{1},0)$ \\
Vector leptoquark & $ V_{\mu} $  & $(\textbf{3},\textbf{1}, \frac{2}{3}) $ \\
$ U(1)_{Y} $ electroweak boson field & $ B_{\mu} $  & $(\textbf{1},\textbf{1}, 0) $ \\
Gluon & $ G_{\mu} $  & $(\textbf{8},\textbf{1}, 0) $ \\
 [1ex] \hline
\end{tabular}
\caption{List of fields in (\ref{2-1}) and their symmetry properties.}
\label{table1}
\end{table}

The fields of (\ref{2-1}) and their symmetry properties has been listed in table \ref{table1}.
In the summation of (\ref{2-1}), for simplicity, we have avoided mixing terms between generations. To keep it simple, we have also assumed the couplings $ g_{L} $, $ g_{R} $, and $ g_{\chi} $ is independent of generations.

\begin{figure}[!htb]
\begin{center}
\includegraphics[scale=0.7]{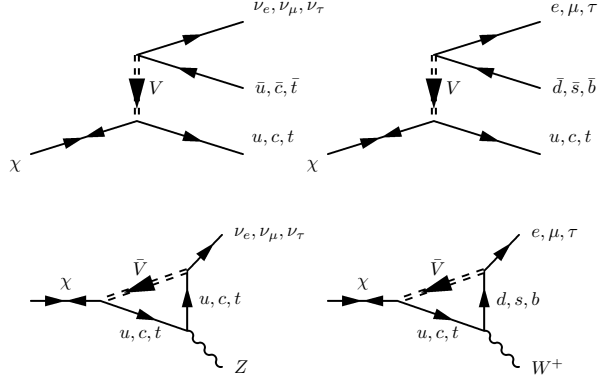}
\end{center}
\caption{Some decay modes of Majorana particle. \label{Decay}}
\end{figure}

In our model Majorana spinor can be a DM candidate if $ M_{\chi} < M_{V} $, otherwise the two-body decay of $ \chi $ to vector leptoquark $ V $ and up-type anti-quarks occurs at tree level and Majorana particle will be unstable.
Even if $ M_{\chi} < M_{V} $, in the case of non-zero couplings $ g_{L} $ and $ g_{R} $, still tree level three-body decay and one-loop induced decay of $ \chi $ can occur (see figure \ref{Decay}). To evade such decays, we impose a discrete $ Z_{2} $ symmetry under which only vector leptoquark and Majorana spinor are odd \cite{Mandal:2018czf}.
In this case, $ g_{L} $ and $ g_{R} $ are zero and Majorana particle can serve as a cosmological stable DM candidate. Moreover, relaxing $ Z_{2} $ symmetry, the constraint on
DM lifetime leads to highly suppressed $ g_{L} $ and $ g_{R} $ for $ {\cal{O}} (1) $ $  g_{\chi} $ values \cite{Garny:2012vt,Arcadi:2013aba,Arcadi:2014dca}.
For the rest of the paper, we assume $ g_{L} $ and $ g_{R} $ are zero. Therefore, in the last line of Lagrangian (\ref{2-1}), the term with $  g_{\chi} $ coupling plays the important role in linking the visible and dark sector to each other.

In unitary gauge, we have $ H = \frac{1}{\sqrt{2}} \begin{pmatrix}
0 \\ h \end{pmatrix} $, and the potential terms (line 2 in (\ref{2-1})) become:
\begin{equation}
- V(h,\phi) = - \frac{1}{4 !} \lambda_{H} h^{4} - \frac{1}{4 !} \lambda_{\phi} \phi^{4} - \frac{1}{2} \lambda_{\phi H} h^{2} \phi^{2}. \label{2-3}
\end{equation}
Vacuum expectation values, $ \langle h \rangle = \nu_{h} $ and $ \langle \phi \rangle = \nu_{\phi} $, correspond to
local minimum of $ V(h,\phi) $. The potential $ V(h,\phi) $ has local minimum if
\begin{align}
& \frac{\partial V(h,\phi)}{\partial h} \bigg\rvert_{\nu_{h},\nu_{\phi}} = \frac{\partial V(h,\phi)}{\partial \phi} \bigg\rvert_{\nu_{h},\nu_{\phi}} = 0, \label{2-4} \\
& \frac{\partial^{2} V(h,\phi)}{\partial h^{2}} \bigg\rvert_{\nu_{h},\nu_{\phi}} > 0 , \label{2-5} \\
& \left( \frac{\partial^{2} V(h,\phi)}{\partial h^{2}} \bigg\rvert_{\nu_{h},\nu_{\phi}} \right) \left( \frac{\partial^{2} V(h,\phi)}{\partial \phi^{2}} \bigg\rvert_{\nu_{h},\nu_{\phi}} \right) - \left( \frac{\partial^{2} V(h,\phi)}{\partial h \, \partial \phi } \bigg\rvert_{\nu_{h},\nu_{\phi}}\right) ^{2} > 0 . \label{2-6}
\end{align}
Eq.~(\ref{2-4}) leads to $ \lambda_{H} \lambda_{\phi} = (3! \lambda_{\phi H})^{2} $ and the following constraint
\begin{equation}
\frac{\nu_{h}}{\nu_{\phi}} = \sqrt{- \frac{3! \lambda_{\phi H}}{\lambda_{H}}}. \label{2-7}
\end{equation}
Vacuum stability, constraints (\ref{2-4}) and (\ref{2-5}), implies that $ \lambda_{H} > 0 $, $ \lambda_{\phi} > 0 $, and $ \lambda_{\phi H} < 0 $. Constraint (\ref{2-7}) defines a stationary line or a local minimum line, known as flat direction, in which $ V (\nu_{h},\nu_{\phi}) = 0 $.
Therefore, the one-loop effective potential dominates along the flat direction. In this direction,
due to one-loop corrections, a small curvature appears with a minimum as the vacuum expectation value $ \nu^{2} = \nu_{h}^{2} + \nu_{\phi}^{2} $ characterized by a RG scale $ \Lambda $.
Therefore, we substitute $ h \rightarrow \nu_{h} + h $ and $ \phi \rightarrow \nu_{\phi} + \phi $ as a result of spontaneous symmetry breaking where $ \nu_{h} = 246 $ GeV.

We define the mass eigenstates $ H_{1} $ and $ H_{2} $ as
\begin{equation}
\begin{pmatrix}
H_{1}\\H_{2}\end{pmatrix}
 =\begin{pmatrix} cos \alpha~~~  -sin \alpha \\sin \alpha  ~~~~~cos \alpha
 \end{pmatrix}\begin{pmatrix}
h \\  \phi
\end{pmatrix}. \label{2-8}
\end{equation}
The scalar field $ H_{2} $ is along the flat direction, thus $ M_{H_{2}} = 0 $ at the tree level, while $ H_{1} $ is perpendicular to the flat direction and we consider it as the SM-like Higgs observed at the LHC with $ M_{H_{1}} = 125 $ GeV. 
We have these constraints following the symmetry breaking:
\begin{align}
& \nu_{\phi} =  \frac{M_{\chi}}{g_{\phi}} , \quad tan \alpha =  \frac{\nu_{h}}{\nu_{\phi}}  , \quad \lambda_{H} =  \frac{3 M_{H_{1}}^{2}}{ \nu_{h}^{2}} cos^{2} \alpha , \quad \lambda_{\phi} =  \frac{3 M_{H_{1}}^{2}}{ \nu_{\phi}^{2}} sin^{2} \alpha , \nonumber  \\
&   \lambda_{\phi H} =  - \frac{ M_{H_{1}}^{2}}{4 \nu_{h} \nu_{\phi} } sin 2 \alpha ,  \quad \lambda_{\phi V} = - \frac{1}{2} \lambda_{H V} \frac{\nu_{h}^{2}}{\nu_{\phi}^{2}} - \frac{M_{V}^{2}}{\nu_{\phi}^{2}},  \label{2-9}
\end{align}
where $ M_{\chi} $ and $ M_{V} $ are the mass of Majorana DM and vector leptoquark after symmetry breaking. In the next sections, to get a more minimal model, we put $ \lambda_{HV} = 0 $. Note that non-zero $ \lambda_{HV} $ does not add any new vertex to our model. Therefore, according to constraints (\ref{2-9}), we consider four free parameters in our model:
\begin{equation}
M_{\chi},M_{V},g_{\chi},g_{\phi} . \nonumber
\end{equation}

As we mentioned before, the scalon field $ H_{2} $ is massless in tree level. However, using Gildener-Weinberg mechanism \cite{Gildener:1976ih},
the radiative corrections give a mass to $ H_{2} $. 

The one-loop effective potential, Along the flat direction, takes the form
\begin{equation}
V_{eff}^{1-loop} = a H_{2}^{4} + b H_{2}^{4} \, \ln \frac{H_{2}^{2}}{\Lambda^{2}}  , \label{2-10}
\end{equation}
where $ a $ and $ b $ are dimensionless constants given by
\begin{align}
& a =  \frac{1}{64 \pi^{2} (\nu_{h}^{2} + \nu_{\phi}^{2})^{2}}  \sum_{k=1}^{n} g_{k}  M_{k}^{4} \ln \frac{M_{k}^{2}}{\nu^{2}}  , \nonumber \\
& b = \frac{1}{64 \pi^{2} (\nu_{h}^{2} + \nu_{\phi}^{2})^{2}} \sum_{k=1}^{n} g_{k}  M_{k}^{4} . \label{2-11}
\end{align}
and $ M_{k} $  ($ g_{k} $) is the  tree-level mass (the internal degrees of freedom) of the particle $ k $. Note that $ g_{k} $
is positive (negative) for bosons (fermions).

\begin{figure}[!htb]
\begin{center}
\includegraphics[scale=0.8]{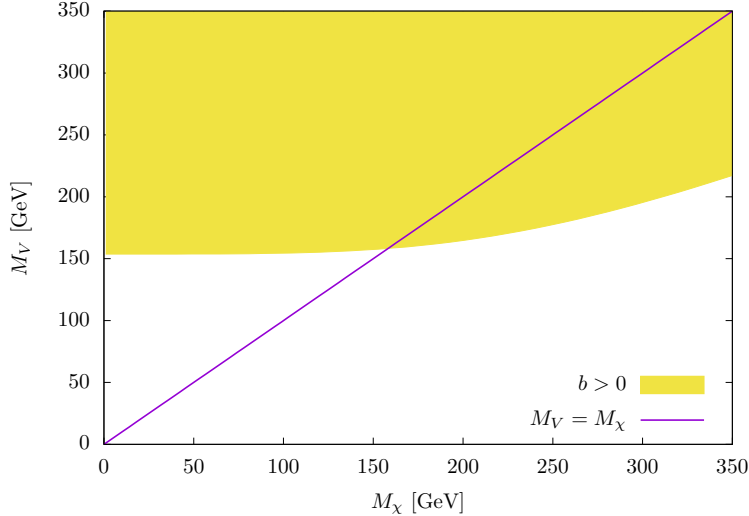}
\end{center}
\caption{Symmetry breaking ocuures if $ b > 0 $. \label{mass}}
\end{figure}

In terms of the one-loop VEV $ \nu $, effective potential along the flat direction is given by
\begin{equation}
V_{eff}^{1-loop} = b H_{2}^{4} \, \left( \ln \frac{H_{2}^{2}}{\nu^{2}} - \frac{1}{2} \right) , \label{2-12}
\end{equation}
and the scalon mass will be
\begin{equation}
M_{H_{2}}^{2} = \frac{d^2 V_{eff}^{1-loop}}{d H_{2}^{2}} \bigg\rvert_{\nu} = 8 b \nu^{2} . \label{2-13}
\end{equation}
According to (\ref{2-11}) and (\ref{2-13}), the mass of scalon can be expressed as
\begin{equation}
M_{H_{2}}^{2} = \frac{1}{8 \pi^{2} (\nu_{h}^{2} + \nu_{\phi}^{2})} \left( M_{H_{1}}^{4} + 6  M_{W}^{4} + 3  M_{Z}^{4} + 18  M_{V}^{4} - 12 M_{t}^{4} - 2 M_{\chi}^{4}  \right) , \label{2-14}
\end{equation}
where $ M_{W} $, $ M_{Z} $ are the masses of W and Z gauge bosons, respectively, and $ M_{t} $ is the mass of top quark.

To get spontaneous symmetry breaking, the minimum of the one-loop potential $ V_{eff}^{1-loop} $ should be negative, thus, $ b $ should be positive. Note that the presence of vector leptoquark is essential to get positive $ b $. Indeed, to get a positive $ b $ we should have $ M_{V} \gtrsim 150 $ GeV (see figure \ref{mass}). According to figure \ref{mass}, for $ M_{\chi} \lesssim 150 $ GeV, the constraint $ M_{\chi} < M_{V} $, which avoids DM decay, is automatically satisfied duo to symmetry breaking condition, i.e., $ b > 0 $. For $ M_{\chi} \gtrsim 150 $ GeV we put aside a part of mass parameter space by hand in order to avoid DM decay.

\section{Relic density} \label{Relic density}
\begin{figure}[!htb]
\begin{center}
\centerline{\hspace{0cm}\epsfig{figure=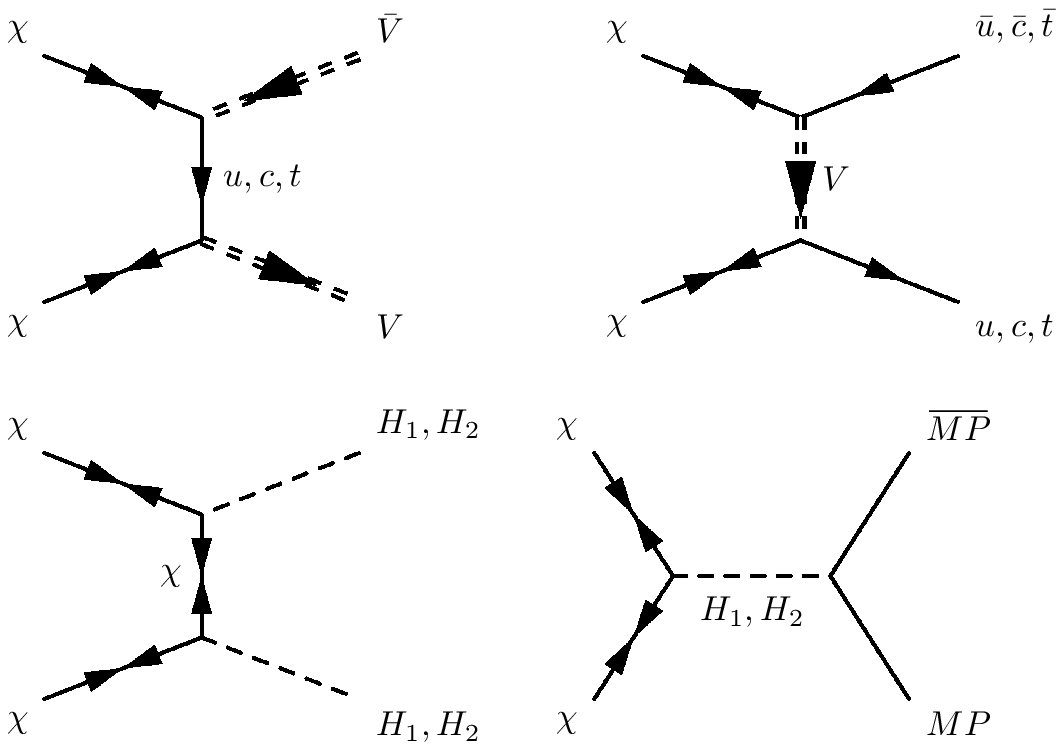,width=7cm}\hspace{0.3cm}\epsfig{figure=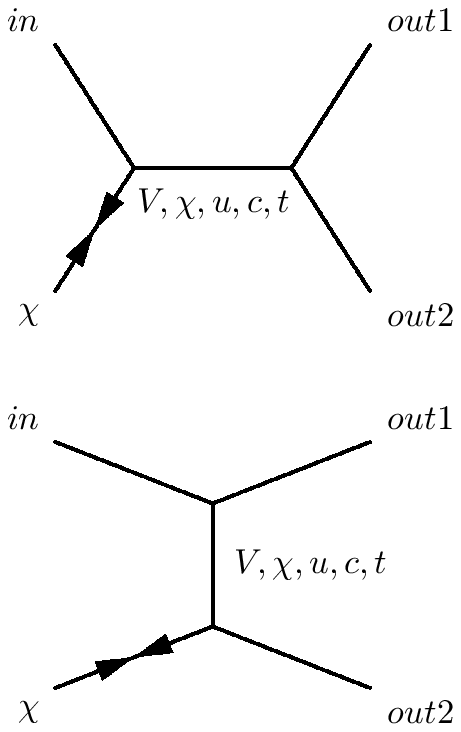,width=3.75cm}}
\centerline{\vspace{-1cm}\hspace{1cm}(a)\hspace{5.5cm}(b)}
\centerline{\vspace{-0.0cm}}
\end{center}
\caption{(a) DM annihilation and (b) coannihilation channels. In this figure MP stands for massive particle.}\label{channels}
\end{figure}

If DM does not interact sufficiently in the early Universe, it will fall out of local thermodynamic equilibrium and it is said to be decoupled. This happens when DM interaction rate drops below the expansion rate of the Universe. To calculate DM relic density one should use Boltzmann equation in which DM annihilation cross sections is needed. Feynman diagrams for all possible DM annihilation channels is depicted in figure \ref{channels} (a). DM annihilates through s-channel in Higgs portal and t-channel in leptoquark portal. Since Majorana particle is its own antiparticle, for every t-channel annihilation there is also a u-channel diagram.
In our model, coannihilation channels also exist (see figure \ref{channels} (b)).

\begin{figure}[!htb]
\begin{center}
\centerline{\hspace{0cm}\epsfig{figure=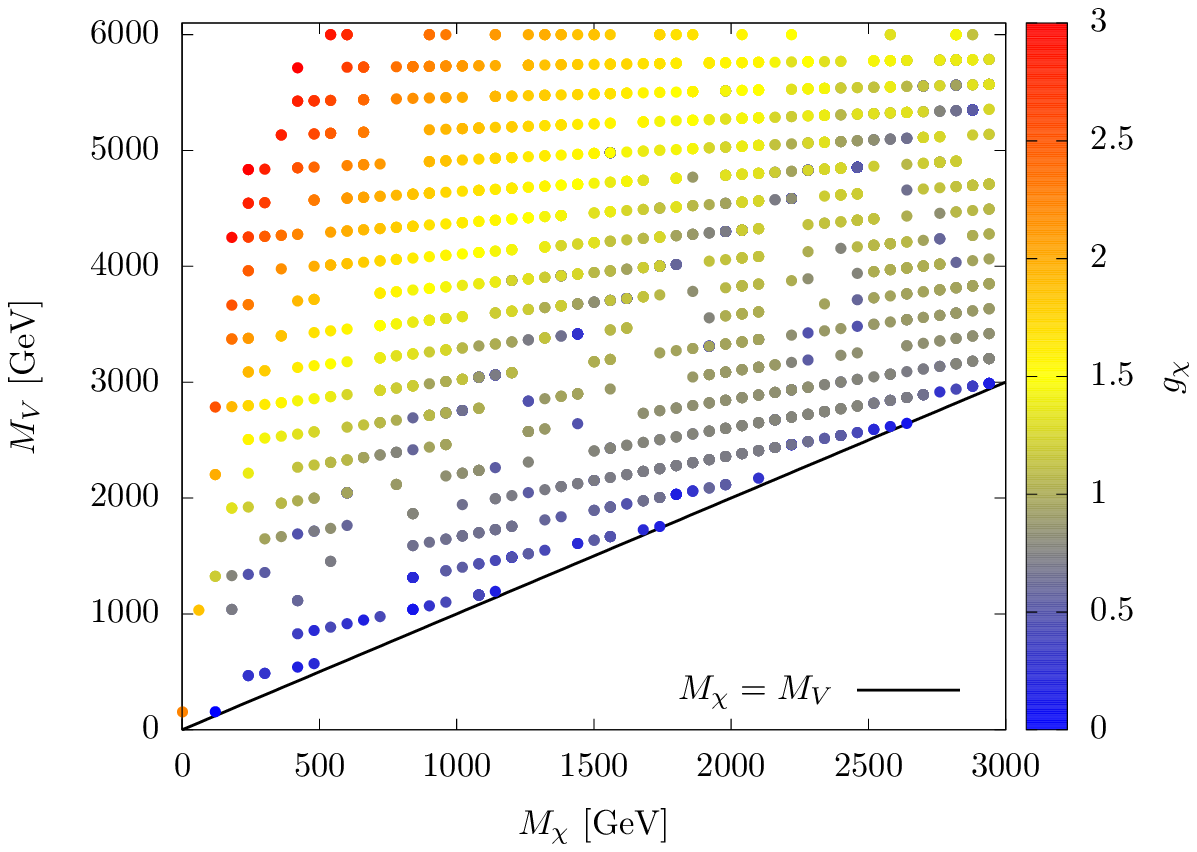,width=7cm}\hspace{0.3cm}\epsfig{figure=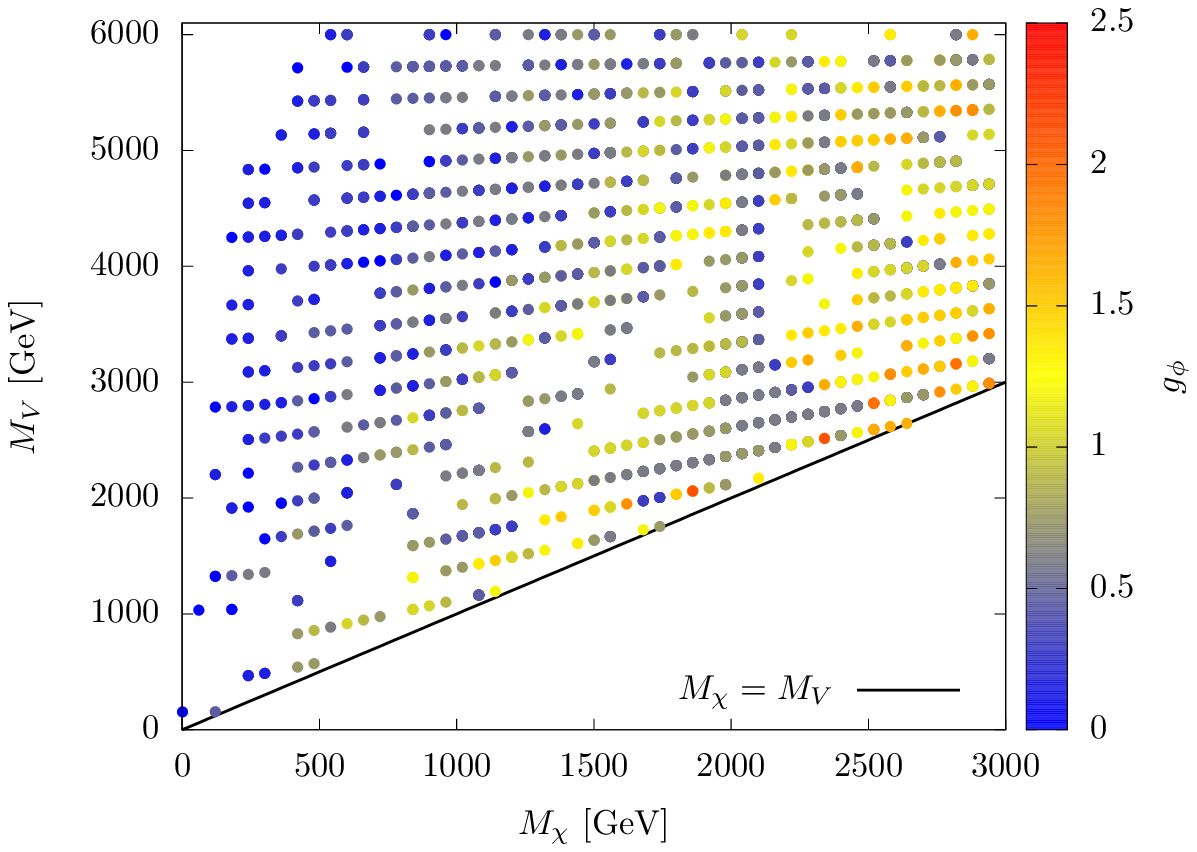,width=7cm}}
\end{center}
\caption{The parameter space of the model constrained by DM relic density as reported by Planck collaboration \cite{Aghanim:2018eyx}.}\label{relic}
\end{figure}

Coannihilation channels are relevant if there is some other particle nearly degenerate in mass with the DM such that it annihilates with DM more efficiently
than DM with itself. In this case, coannihilation channels primarily determines DM relic density. To quantitatively account DM relic
density, one should solve Boltzmann equation \cite{Hooper:2009zm}
\begin{equation}
\frac{d n_{\chi}}{dt} + 3H n_{\chi}= - \langle\sigma_{eff} |v_{rel}|\rangle (n_{\chi}^{2} - n_{\chi,eq}^{2}), \label{3-1}
\end{equation}
where $ n_{\chi} $ is the number density of Majorana DM, $ H $ is the Hubble parameter, and $ \langle\sigma_{eff} |v_{rel}|\rangle $ is the thermally averaged of effective annihilation cross section (multiplied by relative velocity). Effective annihilation $ \sigma_{eff} $ is given by \cite{Hooper:2009zm}
\begin{equation}
\sigma_{eff} = \sum_{i,j} \sigma_{i,j} \frac{g_{i}g_{j}}{g_{eff}^{2}} (1+\Delta_{i})^{3/2} (1+\Delta_{j})^{3/2} e^{-\frac{M_{\chi}}{T}(\Delta_{i}+\Delta_{j})} , \label{3-2}
\end{equation}
where the double sum is over all particle species with $ \sigma_{1,1} $ being Majorana DM annihilation cross section and $ \sigma_{i,j} $ is the cross section for the coannihilation of species $ i $ and $ j $ (or
self-annihilation in the case of $ i=j $) into Standard Model particles. The quantities $ \Delta_{i} = (M_{i}-M_{\chi})/M_{\chi} $ is the fractional mass splittings between the species $ i $ and the Majorana DM and $ g_{eff} = \sum_{i} g_{i} (1+\Delta_{i})^{3/2} e^{-\frac{M_{\chi}}{T}\Delta_{i}} $. In order to calculate Majorana DM relic density including coannihilations channels, we use the public numerical code {\tt micrOMEGAs} \cite{Barducci:2016pcb}. The Lagrangian (\ref{2-1}) has been implemented through {\tt LanHEP} \cite{Semenov:2014rea} package. We use DM relic density ($ \Omega_{DM} h^{2} = 0.120 \pm 0.001 $) reported by Planck \cite{Aghanim:2018eyx} as a
constraint in scanning the four dimensional parameter space of the model. The result is depicted in figure \ref{relic}.

\section{Direct and indirect detection} \label{Direct detection}
\begin{figure}[!htb]
\begin{center}
\includegraphics[scale=0.8]{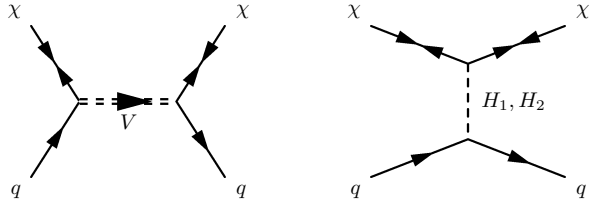}
\end{center}
\caption{Feynman diagrams  responsible for DM-nucleon scattering. \label{Direct}}
\end{figure}

Majorana DM can elastically scatter off the nucleus. The momentum transfer gives rise to a nuclear recoil which might produce a signal in direct detection experiments. In our model, this signal can arise from the Feynman diagrams shown in figure \ref{Direct}.

Both spin-dependent (SD) and spin-independent (SI) DM-nucleon scattering exist in our model. The left (right) diagram of figure \ref{Direct} leads to SD (SI) scattering which can be described by effective axial-vector (scalar) Lagrangian $ {\cal{L}}_{A} = c_{A} \overline{\chi} \gamma_{\mu} \gamma_{5} \chi \overline{u} \gamma^{\mu} \gamma^{5} u  $ ($ {\cal{L}}_{S} = c_{S,q} \overline{\chi} \chi \overline{q} q  $). To obtain $ c_{A} $ and $ c_{
S,q} $ we should integrate out the intermediate particles shown in Feynman diagrams \ref{Direct}. The result is
\begin{equation}
c_{A} = - \frac{g_{\chi}^{2}}{4(M_{V}^{2}-M_{\chi}^{2})}, \quad c_{S,q} = - \frac{m_{q}}{4 \nu_{h}} g_{\phi} \sin 2 \alpha \left( \frac{1}{M^{2}_{H_{1}}} - \frac{1}{M^{2}_{H_{2}}} \right) , \label{4-1}
\end{equation}
Having coefficients $ c_{A} $ and $ c_{
S,q} $, SD and SI DM-nucleon cross sections become \cite{Agrawal:2010fh}
\begin{align}
&\sigma^{SD} = \frac{16 \mu^{2}_{N \chi}}{\pi} c^{2}_{A} (\Delta_{\mu}^{N})^{2} J_{N}(J_{N}+1), \label{4-2} \\
&\sigma^{SI} = \frac{4 \mu^{2}_{N \chi} M^{2}_{
N}}{\pi} \frac{c_{S,q}^{2}}{m_{q}^{2}} f_{N}^{2}, \label{4-3}
\end{align}
where $ \mu_{N \chi} = M_{\chi} M_{N}/(M_{\chi} + M_{N}) $ is the reduced mass of DM and nucleon, $ J_{N} = \frac{1}{2} $ is the angular momentum of the nucleon, $ \Delta_{\mu}^{N} = 0.78 \pm 0.02 $ ($ \Delta_{\mu}^{N} = -0.48 \pm 0.02 $) is the u-quark spin fraction in the proton (neutron) \cite{Mallot:1999qb,Ellis:2000ds}, and $ f_{N} \simeq 0.3 $ parametrizes the Higgs-nucleon coupling.

\begin{figure}[!htb]
\begin{center}
\centerline{\hspace{0cm}\epsfig{figure=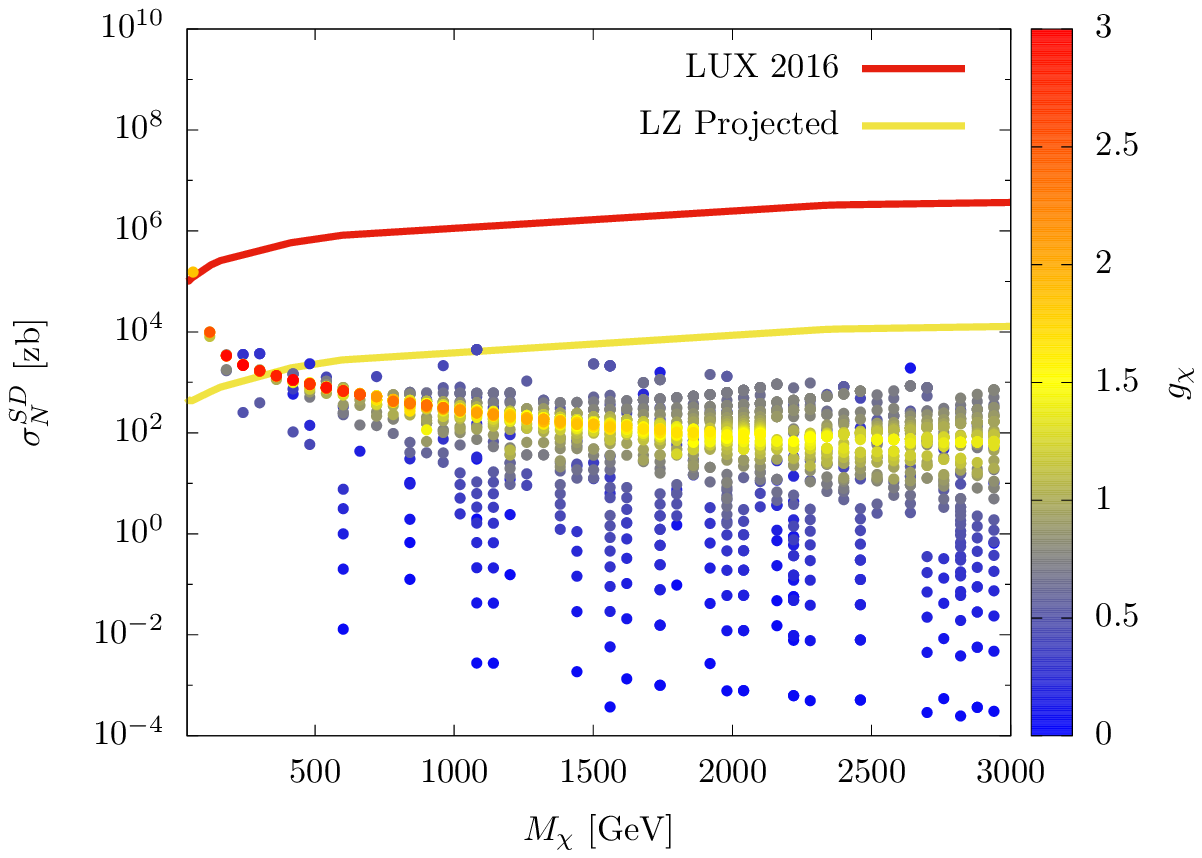,width=7cm}\hspace{0.3cm}\epsfig{figure=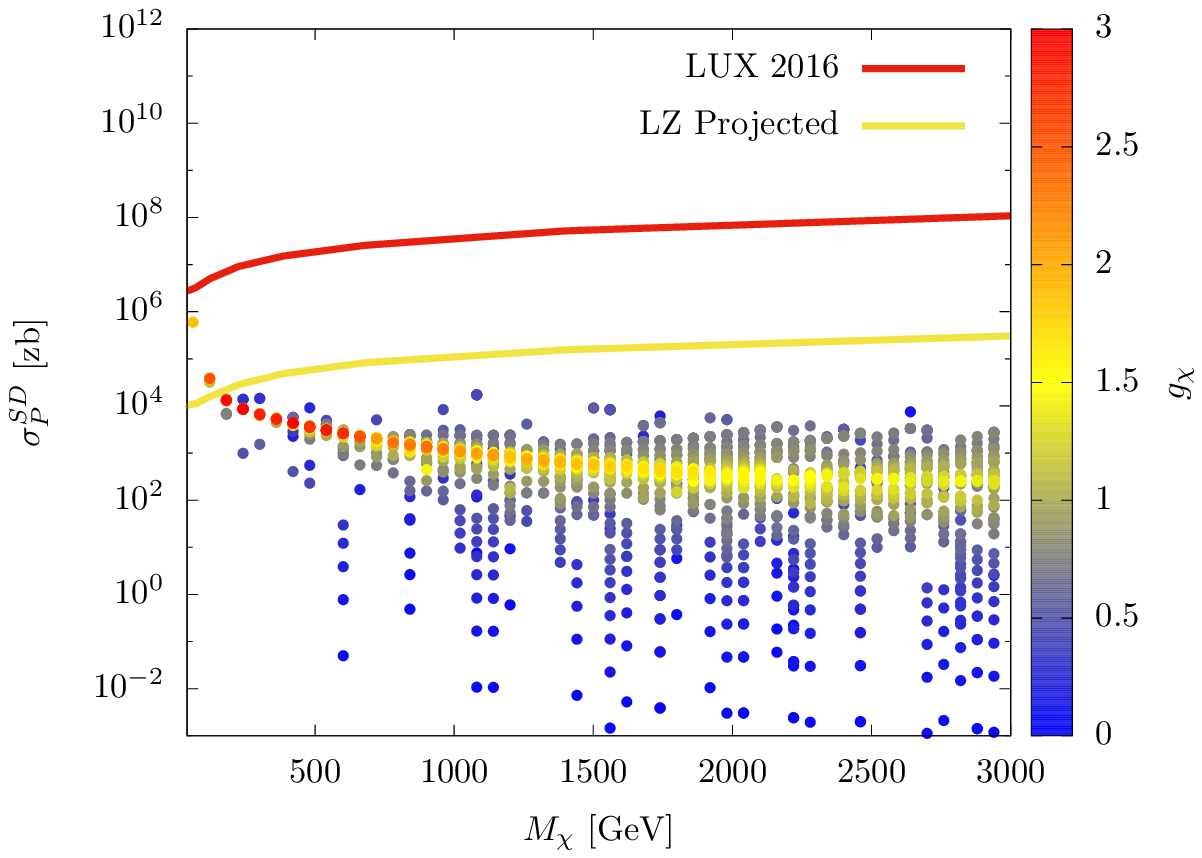,width=7cm}}
\centerline{\vspace{-1cm}\hspace{0.5cm}(a)\hspace{6cm}(b)}
\centerline{\vspace{-0.0cm}}
\end{center}
\caption{The direct detection SD cross section vs DM mass for (a) DM-neutron and (b) DM-proton scattering.}\label{DirectSD}
\end{figure}

\begin{figure}[!htb]
\begin{center}
\centerline{\hspace{0cm}\epsfig{figure=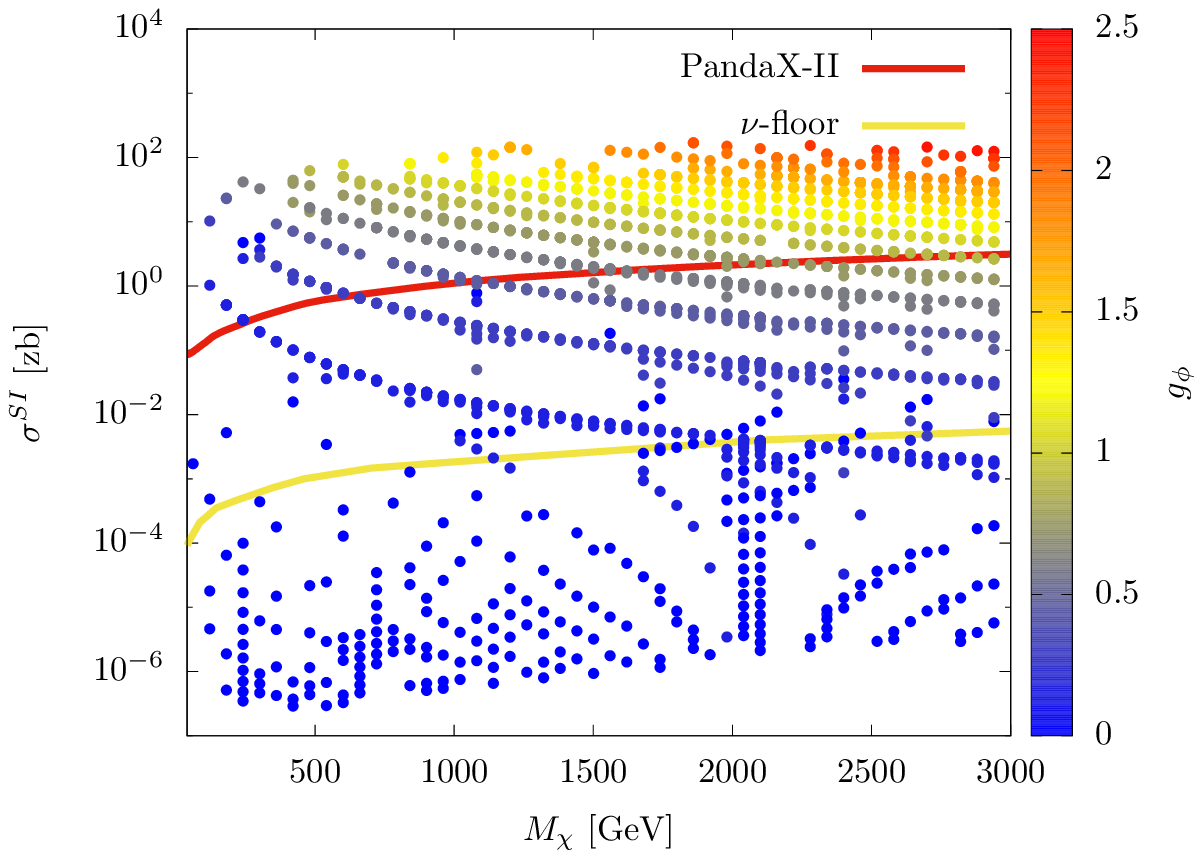,width=7cm}}
\centerline{\vspace{-1cm}\hspace{0.5cm}(a)}
\vspace{1.3cm}{\epsfig{figure=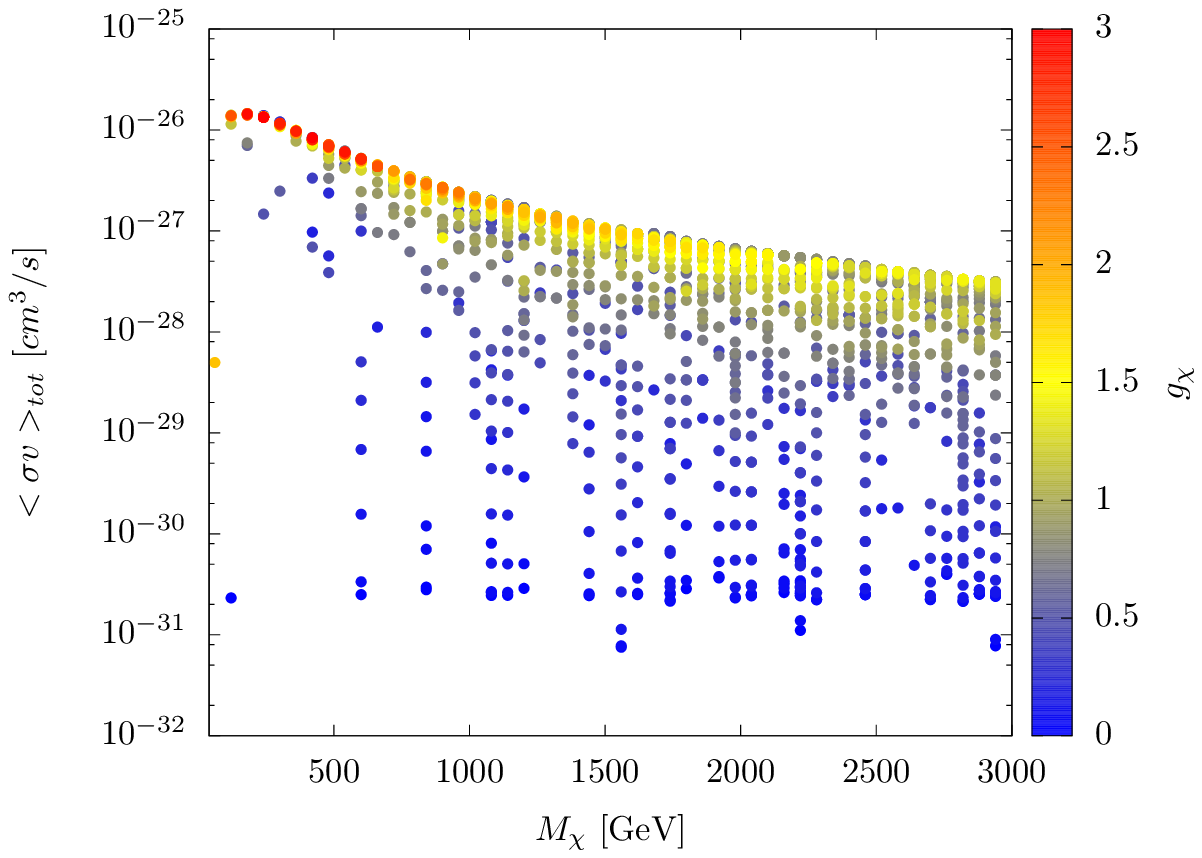,width=7cm}\hspace{0.3cm}\epsfig{figure=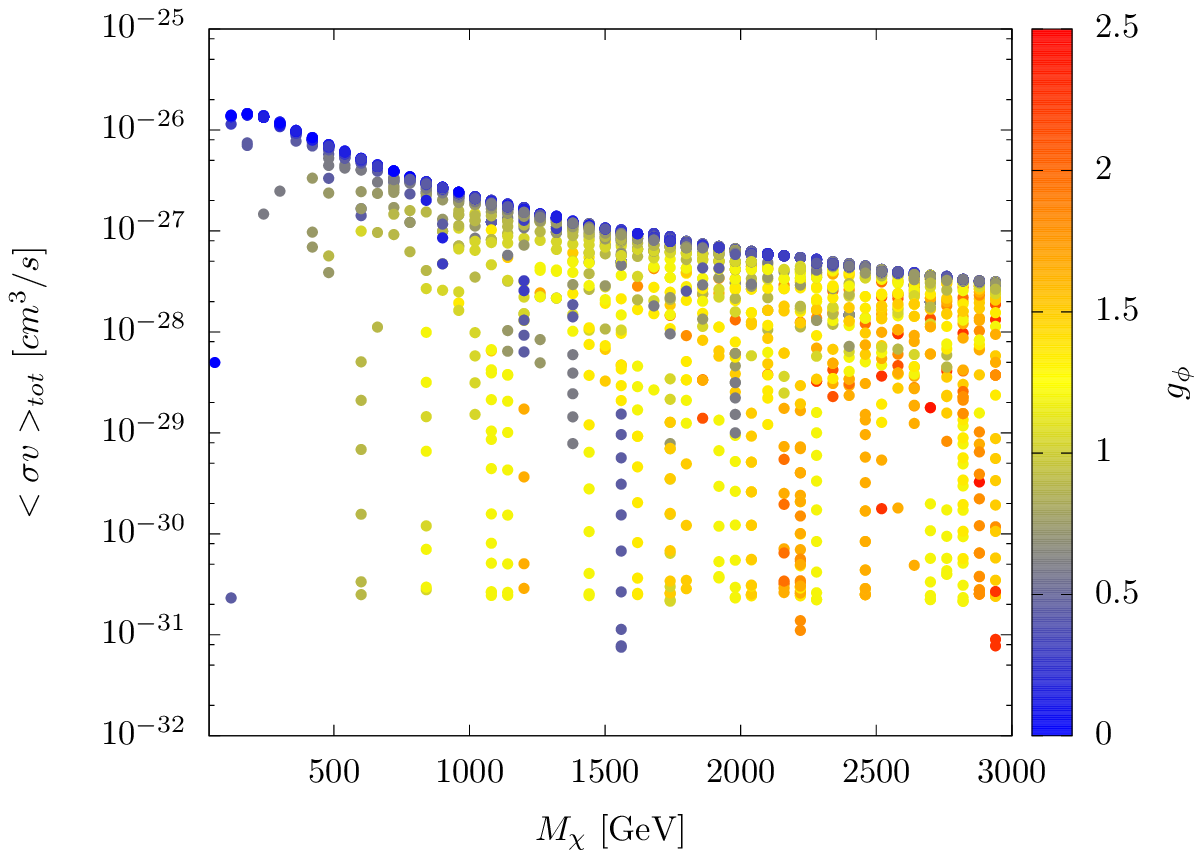,width=7cm}}
\centerline{\vspace{-1cm}\hspace{0.5cm}(b)\hspace{6cm}(c)}
\centerline{\vspace{-0.0cm}}
\end{center}
\caption{(a) SI DM-nucleon cross section. (b) and (c) DM total velocity-averaged annihilation cross section vs DM mass.}\label{DirectIndirect}
\end{figure}

Direct detection experiments put upper limits on SD and SI DM-nucleon cross sections. We calculate these cross sections using {\tt micrOMEGAs} package. In order to constrain the model, we use PandaX-II \cite{Cui:2017nnn} and LUX \cite{Akerib:2016lao} experiments for SI and SD DM-nucleon scattering, respectively. In figure \ref{DirectSD} we have depicted SD DM-nucleon cross section for the parameter space compatible with DM relic density. As it is seen in this figure, the model can evade the upper limit of SD DM-nucleon cross section. However, a small part of the parameter space can be probed by future LZ experiment \cite{Akerib:2015cja}. Unlike the upper limit of SD DM-nucleon cross section, as it is depicted in \ref{DirectIndirect} (a), PandaX-II upper limit of SI DM-nucleon scattering excludes some part of the parameter space already constrained by DM relic density. In this figure, we have also shown the neutrino floor \cite{Billard:2013qya} which  limits the parameter space from below from the irreducible background of coherent neutrino-nucleus scattering. 

\begin{table}
\centering
\begin{tabular}{c c c c c c c c} 
 \hline
$ \# $ & $ \sin \alpha $ & $ g_{\phi} $ & $ g_{\chi} $ & $ M_{\chi} $ (GeV) & $ M_{V} $ (GeV) & $ M_{H_{2}} $ (GeV) & $ \langle \sigma v_{rel} \rangle $ ($ cm^3/s $) \\ [0.5ex] 
 \hline
1 & 0.001 & 0.001 & 3.000 & 240.9  & 4835 & 46.36 & $ 1.336 \times 10^{-26} $ \\
2 & 0.293 & 1.051 & 0.121 & 840.7  & 1039 & 600.8 & $ 1.449 \times 10^{-29} $ \\
3 & 0.070 & 0.450 & 0.001 & 1560  & 4980 & 3413 & $ 7.525 \times 10^{-32} $ \\
 [1ex] \hline
\end{tabular}
\caption{Total DM annihilation cross section in galactic halos for three benchmark points constrained by the measured cosmological DM relic density, i.e., $ \Omega h^2 \simeq 0.12 $.}
\label{table2}
\end{table}

\begin{table}[th]
\begin{center}
\begin{tabular}{|c|c|c|c|c|}
\multicolumn{1}{c}{} & \multicolumn{2}{c}{freeze-out temperature} & \multicolumn{2}{c}{today temperature} \\ \hline
$ \# $& main channels & contribution & main channels & contribution \\ \hline
1 & \makecell{
   $ \chi,\chi \rightarrow t, \overline{t} $ \\
   $ \chi,\chi \rightarrow c, \overline{t} $ \\ 
   $ \chi,\chi \rightarrow t, \overline{c} $ \\ 
   $ \chi,\chi \rightarrow u, \overline{t} $ \\ 
   $ \chi,\chi \rightarrow t, \overline{u} $
}
 & \makecell{
   23\% \\
   15\% \\ 
   15\% \\ 
   15\% \\
   15\%
} & \makecell{
   $ \chi,\chi \rightarrow t, \overline{t} $ \\
   $ \chi,\chi \rightarrow c, \overline{t} $ \\ 
   $ \chi,\chi \rightarrow t, \overline{c} $ \\ 
   $ \chi,\chi \rightarrow u, \overline{t} $ \\ 
   $ \chi,\chi \rightarrow t, \overline{u} $} & \makecell{
   31\% \\
   17\% \\ 
   17\% \\ 
   17\% \\
   17\%}\\ \hline
2 & \makecell{
   $ \chi,\chi \rightarrow H_{2}, H_{2} $
}
 & \makecell{
   84\%
} & \makecell{
   $ \chi,\chi \rightarrow t, \overline{t} $ \\
   $ \chi,\chi \rightarrow c, \overline{t} $ \\ 
   $ \chi,\chi \rightarrow t, \overline{c} $ \\ 
   $ \chi,\chi \rightarrow u, \overline{t} $ \\ 
   $ \chi,\chi \rightarrow t, \overline{u} $
} & \makecell{
   32\% \\
   16\% \\ 
   16\% \\ 
   16\% \\
   16\%}\\ \hline
3 & \makecell{
   $ \chi,\chi \rightarrow W^{+}, W^{-} $ \\
   $ \chi,\chi \rightarrow Z, Z $
}
 & \makecell{
   66\% \\
   33\% 
} & \makecell{
   $ \chi,\chi \rightarrow W^{+}, W^{-} $ \\
   $ \chi,\chi \rightarrow Z, Z $} & \makecell{
   66\% \\
   33\%}\\ \hline
\end{tabular}
\caption{Main DM annihilation channels both at thermal freeze-out and today temperatures. For benchmark point 1, DM annihilates through s-wave leptoquark portal at both temperatures and, therefore, its annihilation cross section is not suppressed, while for benchmark point 3, annihilation channels are p-wave suppressed due to DM annihilating through Higgs portal. For benchmark 2, at freeze-out temperature DM annihilates through Higgs portal, while its annihilation at today Universe is through leptoquark portal. Roughly speaking, one can say DM annihilation cross section is close to freez-out one for large values of $ g_{\chi} $ where DM mostly annihilates via leptoquark portal.}
\label{table3}
\end{center}
\end{table}

Using {\tt micrOMEGAs}, we have also calculated DM total annihilation cross section in the modern Universe for the parameters which are already constrained by DM relic density. The result is depicted in figure \ref{DirectIndirect} (b) and (c). Furthermore, for the benchmark points
shown in table~\ref{table2}, different DM annihilation channels, both at freez-out and today temperatures, are reported in table~\ref{table3}.
Today DM annihilation cross section is relevant in indirect searches for DM which include attempts to detect the gamma rays, positrons, antiprotons, neutrinos, and
other particles that are produced in DM annihilations or decays. Stable DM particles with a thermally averaged annihilation cross section of $ \langle \sigma v_{rel} \rangle \sim {\cal{O}}(10^{-26})
\, cm^{3}/s $ is predicted to freeze out of thermal equilibrium with an abundance equal to the measured
cosmological density of DM. Indirect searches for DM, especially gamma ray and
cosmic ray searches for DM annihilation products, have recently become sensitive to this
benchmark cross section for masses up to around the weak scale, i.e., ${\cal{O}} (10^2) $ GeV. However,
our model generally evades indirect detection constraints.
As figure \ref{DirectIndirect} (b) and (c) shows, Majorana DM mostly annihilates with a smaller cross section than $ \langle \sigma v_{rel} \rangle \sim {\cal{O}}(10^{-26})
\, cm^{3}/s $ in the universe today. This is because of some velocity dependent DM annihilation cross sections. Velocity dependence of $ \langle \sigma v_{rel} \rangle $ can have
significant effects on the late-time DM annihilation while leaving freeze-out largely unaffected. In appendix \ref{Appendix} some DM annihilation cross sections are written as a Taylor series expansion in powers of $ v_{rel}^{2} $. In this expansion,
p-wave amplitudes only contribute to the $ v_{rel}^{2} $ and higher order terms, while s-wave annihilation amplitudes contribute to all orders. In our model, the s-wave annihilation of Majorana DM to
SM products via Higgs portal is absent. Therefore, these annihilation channels are p-wave
suppressed and DM annihilation cross section is only large
in vector leptoquark portal with large $ g_{\chi} $.
The velocities of DM particles today are around $ v_{rel}\sim10^{-3} $~c, while it is $ v_{rel}\sim0.3 $~c at the temperature of thermal freeze-out. Therefore, we expect the current annihilation rate of a p-wave suppressed DM candidate to be suppressed by a factor of $ \left( \frac{10^{-3}}{0.3}\right)^{2} \sim 10^{-5} $.

\section{Conclusion} \label{Conclusion}
In this paper we discussed a classically accidental scale-invariant extension of SM containing a Majorana DM candidate. DM interacts with SM via two portals, namely, Higgs and vector leptoquark portals.
In the leptoquark sector, we have assumed the DM couples to all generations of up-type quarks with equal coupling. We have also avoid interactions which lead to DM decay by constraining $ M_{\chi} < M_{V} $ and $ g_{L}=g_{R}=0 $. To get a minimal theory, we further assume $ \lambda_{HV} = 0 $. Therefore, we left with four independent parameters which we choose $ M_{\chi} $, $ M_{V} $, $ g_{\chi} $, and $ g_{\phi} $. To put a constraint on these parameters we used Planck data for DM relic density, and LUX and PandaX-II upper bounds for SD and SI DM-nucleon cross sections, respectively.
The parameter space constrained by DM relic density evade SD DM-nucleon cross section upper limit, however SI DM-nucleon cross section excludes some part of the parameter space. We have also shown that our model generally can evade indirect detection constraints because of domination of p-wave DM annihilation cross section in Higgs portal. Therefore, for the parameter space which is already constrained by DM relic density, the annihilation cross section is not large enough to give a signal in indirect detection experiments.

Finally, collider searches for vector leptoquarks via pair and/or single production also impose bound on the leptoquark mass.
CMS 13 TeV data \cite{Sirunyan:2017yrk} excludes $ M_{V} \lesssim 1 $ TeV \cite{DiLuzio:2017chi}. Furthermore, SM Higgs have admixtures
of the scalon which can also bound the parameter space. LHC constraint on the mixing angle between SM Higgs and scalon is $ \sin \alpha \lesssim 0.44 $ \cite{Farzinnia:2013pga,Farzinnia:2014xia}. This bound is compatible with parameter space satisfying DM relic density.

\appendix
\section{DM annihilation cross sections} \label{Appendix}
The leading order s-wave annihilation cross sections of Majorana DM through leptoquark portal (see figure \ref{channels} (a) top right) expanded in powers of $ v_{rel}^{2} $ are given by
\begin{equation}
\sigma v_{rel} (\chi , \chi \rightarrow q_{u} , \overline{q}_{u}) \approx \frac{3 g_{\chi }^4 M_{q_{u}}^2 \sqrt{M_{\chi }^2-M_{q_{u}}^2} \left(-M_{q_{u}}^2+2 M_V^2+M_{\chi }^2\right)^2}{32 \pi  M_V^4 M_{\chi } \left(-M_{q_{u}}^2+M_V^2+M_{\chi }^2\right)^2}, 
\end{equation}
where $ q_{u} $ represents up-type quarks $ \lbrace u,c,t \rbrace $.

The leading order suppressed p-wave annihilation cross sections of Majorana DM through s-channel Higgs portal (see figure \ref{channels} (a) bottom right) expanded in powers of $ v_{rel}^{2} $ are given by
\begin{align}
& \sigma v_{rel} (\chi , \chi \rightarrow \ell , \overline{\ell}) \approx  \frac{n_{c} e^2 g_{\phi }^2 \sin^{2} (2 \alpha) M_{\ell}^2 \left(M_{H_{1}}^2-M_{H_{2}}^2\right)^2 \left(M_{\chi }^2-M_{\ell}^2\right)^{3/2}}{128 \pi  \sin^{2} (\theta_{W}) M_W^2 M_{\chi } \left(M_{H_{1}}^2-4 M_{\chi }^2\right)^2 \left( M_{H_{2}}^2-4 M_{\chi}^2 \right)^2} v_{rel}^{2} , \\
&\sigma v_{rel} (\chi , \chi \rightarrow {\cal{V}} , \overline{{\cal{V}}}) \approx \frac{n_{e} e^2 g_{\phi }^2 \sin^{2} (2 \alpha) \left(M_{H_{1}}^2-M_{H_{2}}^2\right)^2 \sqrt{M_{\chi }^2-M_{\cal{V}}^2}}{512 \pi  \sin^{2} (\theta_{W}) M_W^2 M_{\chi } \left(M_{H_{1}}^2-4 M_{\chi }^2\right)^2 \left( M_{H_{2}}^2-4 M_{\chi}^2 \right)^2} \nonumber \\
& \qquad \qquad \qquad \times \left(4 M_{\chi }^4-4 M_{\chi }^2 M_{\cal{V}}^2+3 M_{\cal{V}}^4\right) v_{rel}^2 , \\
&\sigma v_{rel} (\chi , \chi \rightarrow {\cal{H}} , {\cal{H}}) \approx \frac{g_{\phi }^2 \sqrt{M_{\chi }^2-M_{{\cal{H}}}^2}}{128 \pi  e^2 M_{\chi } \left(M_{H_{1}}^2-4 M_{\chi }^2\right)^2 \left( M_{H_{2}}^2-4 M_{\chi}^2 \right)^2} \nonumber \\
& \qquad \qquad \qquad \times \left(4 M_{\chi }^2 (A_{{\cal{H}}} \sin \alpha-B_{{\cal{H}}} \cos \alpha)-A_{{\cal{H}}} \sin \alpha M_{H_{2}}^2+B_{{\cal{H}}} \cos \alpha M_{H_{1}}^2\right)^2  v_{rel}^2 ,
\end{align}
where $ \ell $, $ {\cal{V}} $, and $ {\cal{H}} $ represent massive fermions, gauge bosons, and scalars, respectively. In these formulas $ \theta_{W} $ is the Weinberg angle, $ n_c=3 $ ($ n_c=1 $) for quarks (leptons), $ n_e=2 $ ($ n_e=1 $) for charged (neutral) gauge bosons, and 
\begin{align}
A_{H_1}=&
2 \cos^3 (\alpha)  \sin (\theta_{W})  \lambda _H M_W-6 \cos^2 (\alpha)  e \, \nu _2 \sin (\alpha)  \lambda _{\phi H}
\nonumber \\
&+12 \cos (\alpha)  \sin^2 (\alpha)  \sin (\theta_{W})  \lambda _{\phi H} M_W-e \, \nu _2 \sin^3 (\alpha)  \lambda _{\phi }, \\
B_{H_1}=&
2 \cos^2 (\alpha)  \sin (\alpha)  \sin (\theta_{W})  \lambda _H M_W+2 \cos (\alpha)  e \, \nu _2 \lambda _{\phi H}+\cos (\alpha)  e \, \nu _2 \sin^2 (\alpha)  \lambda _{\phi }
\nonumber \\
&-6 \cos (\alpha)  e \, \nu _2 \sin^2 (\alpha)  \lambda _{\phi H}+12 \sin^3 (\alpha)  \sin (\theta_{W})  \lambda _{\phi H} M_W-8 \sin (\alpha)  \sin (\theta_{W})  \lambda _{\phi H} M_W, \\
A_{H_2}=&
-\cos^2 (\alpha)  e \, \nu _2 \sin (\alpha)  \lambda _{\phi }+2 \cos (\alpha)  \sin^2 (\alpha)  \sin (\theta_{W})  \lambda _H M_W
\nonumber \\
&-12 \cos (\alpha)  \sin^2 (\alpha)  \sin (\theta_{W})  \lambda _{\phi H} M_W+4 \cos (\alpha)  \sin (\theta_{W})  \lambda _{\phi H} M_W
\nonumber \\
&-6 e \, \nu _2 \sin^3 (\alpha)  \lambda _{\phi H}+4 e \, \nu _2 \sin (\alpha)  \lambda _{\phi H}, \\
B_{H_2}=&
\cos^3 (\alpha)  e \, \nu _2 \lambda _{\phi }+12 \cos^2 (\alpha)  \sin (\alpha)  \sin (\theta_{W})  \lambda _{\phi H} M_W+6 \cos (\alpha)  e \, \nu _2 \sin^2 (\alpha)  \lambda _{\phi H}
\nonumber \\
&+2 \sin^3 (\alpha)  \sin (\theta_{W})  \lambda _H M_W .
\end{align}

Finally, the leading order suppressed p-wave annihilation cross sections of Majorana DM through t-channel Higgs portal (see figure \ref{channels} (a) bottom left) expanded in powers of $ v_{rel}^{2} $ is given by
\begin{equation}
\sigma v_{rel} (\chi , \chi \rightarrow {\cal{H}} , {\cal{H}}) \approx 
\frac{n_{{\cal{H}}}^4 g_{\phi }^4 \, M_{\chi } \sqrt{M_{\chi }^2-M_{{\cal{H}}}^2} \left(-8 M_{{\cal{H}}}^2 M_{\chi }^2+2 M_{{\cal{H}}}^4+9 M_{\chi }^4\right)}{24 \pi  \left(M_{{\cal{H}}}^2-2 M_{\chi }^2\right)^4}  v_{rel}^2
\end{equation}
where $ n_{H_{1}} = \sin \alpha $ and $ n_{H_{2}} = \cos \alpha $.

\section*{Acknowledgment}
This work is supported financially by the Young Researchers and Elite Club of Islamshahr Branch of Islamic Azad University.

\providecommand{\href}[2]{#2}\begingroup\raggedright\endgroup

\end{document}